\documentstyle[prl,aps,multicol,epsfig]{revtex}
\draft
\begin{document}
\date{\today}

\wideabs{
\title{ Coexistence of dimerization and
long-range magnetic order in the frustrated spin chain system
LiCu$_{2}$O$_{2}$: inelastic light scattering study }
\author{K.-Y. Choi,$^{1}$ S. A. Zvyagin,$^{2}$ G. Cao,$^{3}$ and P.
Lemmens$^{1,4}$}
\address{$^1$ 2. Physikalisches Institut, RWTH Aachen, 52056 Aachen,
Germany}
\address{$^2$ National High Magnetic Field Laboratory, 1800 East
Paul Dirac Drive, Tallahassee, Florida 32310, USA}
\address{$^3$ Department of Physics and Astronomy, University of
Kentucky, Lexington, Kentucky 40506, USA}
\address{$^4$ MPI for Solid State Research, 70569 Stuttgart, Germany}
\maketitle

\begin{abstract}
Raman scattering studies of the frustrated spin chain system
LiCu$_{2}$O$_{2}$ are reported. Two transitions into a
magnetically ordered phase (taken place at temperatures $\sim$ 9 K
and $\sim$ 24 K) have been confirmed from the analysis of optical
properties of the samples. Interestingly, two different magnetic
excitations, seen at 100 and 110 cm$^{-1}$ in the magnetically
ordered phase superimpose each other independently, indicating a
coherent coexistence of long-range magnetic order and
dimerization. The observed phenomenon is attributed to
magnetostructural peculiarities of LiCu$_{2}$O$_{2}$ leading to
the intrinsic presence of nonmagnetic impurities  on a nanometer
scale. Furthermore, magnetic impurities play a significant role in
driving the transition from an incommensurate state to a N\'{e}el
ordered one at 9 K.

\end{abstract}


}

\narrowtext

Theoretical and experimental investigations of low-dimensional
spin systems with frustration and dimerization have been boosted
by the discovery of the inorganic spin-Peierls system
CuGeO$_3$.\cite{Hase93} Such a frustrated spin chain system shows
a rich phase diagram, including both gapped and gapless
phases\cite{Lemmens03}, as well as  disorder-induced long-range
ordering, coexisting with a dimerized state.\cite{Regnault95}
Variational calculations suggest that the resonating-valence-bond
character of spin correlations at short distances can be
responsible for the enhancement of antiferromagnetic (AFM)
correlations near vacancies\cite{Martins97,Sorensen98}, that
eventually results in long-range AFM order, locally coexisting
with the disordered dimerized phase. Such a coexistence was
observed in some other doped systems with reduced dimensionality
(for instance, in highly hole-doped chain material
Sr$_{0.73}$CuO$_2$~\cite{Meijer99}) and appears to be a
fundamental property of  low-dimensional quantum spin systems.

LiCu$_{2}$O$_{2}$ can be regarded as a realization of a $S$=1/2
spin chain with competing nearest- and next-nearest-neighbor
interactions.\cite{Berger92,Roessli01,Vorotynov98,Zvyagin02,Masuda03}
This compound has an orthorhombic crystal structure of a space
group Pnma with the lattice parameters $a$=5.72~\AA, $b$=2.86~\AA
\, and $c$=12.4~\AA.\cite{Berger92} There are monovalent and
bivalent copper ions in the unit cell. Magnetic Cu$^{2+}$ ions
form a double-chain along the $b$ axis which is separated from
each other by both Li ions and planes with nonmagnetic Cu$^{+}$
ions. The two separate Cu chains within the double-chain structure
are coupled via a 90$^{\circ}$ oxygen bond along the $c$ axis. At
elevated temperatures high-field electron spin resonance (ESR)
gives evidence for a spin singlet state with spin gap of
$\Delta\sim 72$ K.\cite{Zvyagin02} Interestingly, upon cooling  a
spin singlet state transits into a long-range ordered state with
helimagnetic structure at $T_{c_1}\sim 24$ K.\cite{Masuda03} Some
bulk measurements point to the presence of a second low
temperature transition with a collinear AFM structure at
$T_{c_2}\sim 9$ K.\cite{Zvyagin02} Both transitions are attributed
to an intrinsic non-stoichiometry and the effect of nonmagnetic
and/or magnetic impurities.\cite{Zvyagin02,Masuda03} However, the
exact origin is not yet clear. In addition, with decreasing
temperature a reduction of orthorhombic strain has been
observed.\cite{Roessli01} These magnetostructural peculiarities of
LiCu$_{2}$O$_{2}$ provide a good opportunity to investigate the
influence of a variety of magnetic and nonmagnetic impurities on
magnetic property of a frustrated spin chain system.

Raman spectroscopy has proven to be an extremely powerful
technique to probe magnetic excitations and spin-lattice
interactions in a low-dimensional spin system with unprecedented
precision.\cite{Lemmens03} The main motivation of this work is
using Raman spectroscopy technique to study the nature of  ground
state and the magnetostructural peculiarities in quantum AFM
LiCu$_{2}$O$_{2}$. In the following we will show the coherent
coexistence of long-range ordered states with dimerized state
induced by  nano-scale defects. Further, we propose a possible
mechanism for a disorder-induced long-range ordering in the
studied system.

The single crystals  were grown using a self-flux method and
characterized by a microstructural analysis and thermodynamic
measurements as described in Ref.~\cite{Zvyagin02}. The single
crystals are microscopically twinned and contain LiCuO-impurity
phase. Raman spectra were measured in a quasi-backscattering
geometry with the excitation line $\lambda= 514.5$ of an Ar$^{+}$
laser with the power of 10 mW and were analyzed by a DILOR-XY
spectrometer and a nitrogen cooled charge-coupled device detector.

Figure~1 displays Raman spectra in parallel ($xx$) and crossed
($xy$) polarizations  at 3 K as well as at 5 K and room
temperature in $xx$ polarization. Here the $x$ axis is an
arbitrary direction in the $ab$ plane. Raman spectra in $xy$
polarization exhibit the same behavior as  those in $xx$
polarization with weaker intensity due to a twinning of the single
crystal.\cite{Zvyagin02} Thus, the observed spectra can be
regarded as an average of all $ab$ plane polarizations.
Subtracting the acoustic (B$_{1u}$ + B$_{2u}$ + B$_{3u}$) modes
the factor group analysis of the space group Pnma yields the
following Raman- and infrared-active modes; 10 A$_g$(aa,bb,cc) + 5
B$_{1g}$(ab) + 10 B$_{2g}$(ac) + 5 B$_{3g}$(bc) + 5 A$_{u}$ + 9
B$_{1u}$ + 4 B$_{2u}$ + 9 B$_{3u}$. At room temperature we observe
12 Raman-active modes out of the maximally expected 15 modes in
$ab$ plane polarizations. A broad band extending from 850 to 1300
cm$^{-1}$ is an overtone feature of the first order signals
between 400 and 650 cm$^{-1}$. This might be due to strong
anharmonic lattice interactions and/or resonant scattering.
\begin{figure}[th]
  \begin{center}
   \setlength{\unitlength}{1cm}
    \includegraphics[width= 7cm, clip]{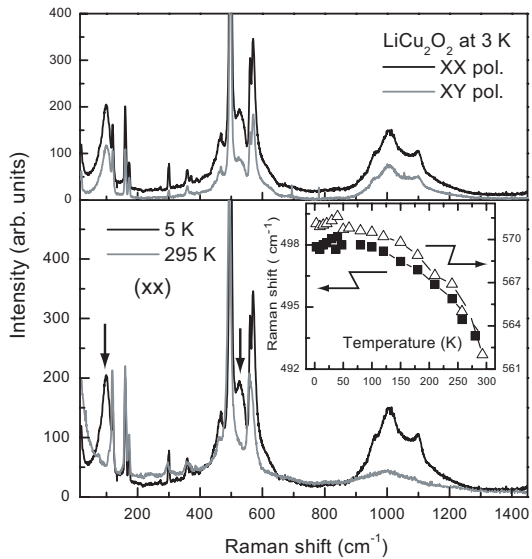}
\end{center}
\caption{Raman spectra of LiCu$_{2}$O$_{2}$ in parallel (xx) and
crossed  (xy) polarizations  at 3 K (upper panel) as well as  at
low and room temperature in parallel polarization (lower panel).
The arrows indicate additional phonon and magnetic signals which
appear at low temperature. Inset: Temperature dependence of the
494- and 562- cm$^{-1}$ phonon modes.}\label{LiCuOf1}
\end{figure}

With decreasing temperature both new phonon modes and magnetic
continua show up in an intriguing manner.  First, we focus on
phonon anomalies which reflect changes of the local symmetry and
bond strengths. Upon cooling new phonon modes at 475 cm$^{-1}$ and
518 cm$^{-1}$ appear below 55 K and 200 K, respectively. The
former temperature corresponds to roughly the exchange constant of
$J\sim 66$ K~\cite{Masuda03} while the latter one is related to
the appearance of the spin gap.\cite{Zvyagin02} Note here that the
number of the observed 14 modes in the low temperature region is
still smaller than the predicted one (15 modes), suggestive of no
reduction of  crystal symmetry at low temperatures. In addition,
almost all phonon modes become sharp and intense while the Cu-O
bending and stretching modes show significant changes as the inset
of Fig.~1 displays. Upon cooling the 494-cm$^{-1}$ and the
562-cm$^{-1}$ modes undergo an appreciable hardening of 5-8
cm$^{-1}$ and then saturate around $T=55$ K. Upon further cooling
they soften slightly while showing a tiny jump at 50 K.
Noticeably, a softening takes place around temperature of the
exchange constant $J\sim 66$ K.\cite{Masuda03} Thus, a softening
below 50 K can be attributed to a renormalization of phonon energy
via spin-phonon coupling, similar to the alternating spin chain
system (VO)$_2$P$_2$O$_7$.\cite{Grove00,Kuhlmann02} The smallness
of the observed softening is related to the big difference in
energy scale between $J\sim 66$ K and the optical phonon energy of
$\sim 700 $ K which mediates the exchange paths along the
double-chain. High-resolution x-ray diffraction
measurements\cite{Roessli01} show a substantial increase of
orthorhombic strain $(a-b)/(a+b)$ upon heating. Transition metal
oxides may show a decrease of the orthorohombicity with increasing
temperature, since thermally activated lattice vibrations reduce
the strain. Normally, phonon frequencies shift to higher energy,
linewidths broaden, and integrated intensity of some phonon modes
decreases. The studied system follows the behavior expected for
usual systems despite the opposite behavior of the orthorhombicity
as a function of temperature. To clarify the seemingly
inconsistency between local and bulk properties detailed studies
of structure as a function of temperature are needed.

We will turn now to distinctive features of magnetic excitations
observed at low temperatures and low frequencies. As Fig.~2
displays, at 3 K a broad asymmetric continuum around 100 cm$^{-1}$
is seen. With increasing temperature the continuum shifts to lower
energy while damping increases. At the same time, a quasielastic
scattering response and a weak continuum around 110 cm$^{-1}$
develop. Although the crystals are twinned, the dynamics of
low-dimensional spin system can be well resolved by Raman
spectroscopy. This is because magnetic Raman scattering of
low-dimensional systems contributes selectively to the
polarization direction in which the incident and scattered light
is parallel to the dominant exchange paths. Thus, twinning does
not add an essential difficulty to interpreting magnetic signals.
To clarify the relation between the magnetic excitations and the
structure of magnetic ordering observed in the bulk material we
will distinguish three phases following Ref.~\cite{Zvyagin02}; I:
magnetic ordered phase (T $<9$ K, presumably with collinear AFM
structure), II: helimagnetic ordered phase~\cite{helim} (9 K $<T<$
23 K), and III: dimerized phase (T $>23$ K).

In phase I the broad continuum extending from 40 cm$^{-1}$ to 130
cm$^{-1}$ is observed. This continuum persists as a wing feature
of quasielastic scattering up to T=21 K (2.3 T$_{c_2}$). This
feature is typical for two-magnon (2M) scattering originating from
double spin-flip processes via the exchange mechanism in
antiferromagnets with collinear structure.\cite{fleury} Therefore,
we identify phase I to be N\'{e}el ordered  with T$_N$=T$_{c_2}=9$
K.

The evolution of the 2M spectrum reflects mainly the temperature
dependence of short-wavelength magnon energies and
lifetimes.\cite{cottam} Thus, the persistence of 2M scattering to
several T$_N$ can be interpreted in terms of the presence of
short-range magnetic fluctuations damped by thermal fluctuations.
In Fig.~3 its temperature dependence of the normalized 2M
frequency and the full width at half-maximum is presented together
with higher dimensional results.\cite{cottam,fleury2} Magnon-pair
energies of LiCu$_2$O$_2$ (1D S=1/2) are renormalized only by 3\%
at T$_N$. In contrast, the magnon-pair spectral weight is
renormalized by 25 \% at $T_{N}$ for 3D S=1/2 systems and by less
than 5\% for 2D S=1 systems.\cite{cottam,fleury2} The damping does
scarcely take place at T$_N$ for LiCu$_2$O$_2$. However, it
strongly increases as the dimension and spin number increase as
the right panel of Fig.~3 displays. The higher dimensionality and
spin number are, the larger are changes of spectral weights at an
energy scale comparable  to the N\'{e}el temperature. This is
related to the fact that  the N\'{e}el temperature is not an
appropriate energy scale for magnetic excitations in
low-dimensional systems. Compared to higher dimensional systems,
the robustness of spin-fluctuation dynamics at the energy scale of
T$_N$ confirms the low-dimensional character of the studied
system.

\begin{figure}[th]
   \begin{center}
    \setlength{\unitlength}{1cm}
    \includegraphics[width= 8cm, clip]{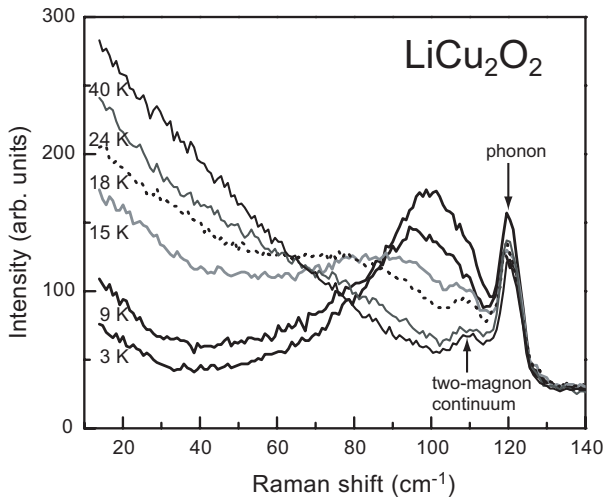}
\end{center}
\caption{ Low frequency Raman scattering of LiCu$_{2}$O$_{2}$.
Quasielastic scattering and two different kinds of two-magnon
continua have been observed. }\label{LiCuOf3}
\end{figure}

The frequency of  the 2M peak  allows an estimate of the
unrenormalized AFM exchange constant between copper spins. The
exact determination of the exchange constant in the studied system
is impossible because no  theoretical calculation for a
double-chain system is known. However, one can make a reasonable
estimation using not too stringent assumptions. Because of
frustrations, the magnetic behavior of LiCu$_{2}$O$_{2}$ lies
between 1D and  2D antiferromagnet. In the 2D case, the peak
energy of $2.7 J$ imposes a lower boundary of the exchange
constant.\cite{Lyons88} In frustrated chain systems the one-magnon
dispersion is given by $\omega(k)=J(1-\alpha cos(kd)/2)$ with the
frustration parameter $\alpha$ and the chain repeat vector
$d$.\cite{Tennant03} In the noninteracting case 2M scattering is
given by twice the magnon density of states. Thus, the peak energy
corresponds to $J(2+\alpha)$. If we assume a renormalization of
the peak energy due to magnon-magnon interactions by the order of
$J\alpha$, then an upper boundary of the exchange constant is
roughly given by $2J$. To conclude, one obtains 52 K $ < J < $ 70
K from the peak energy of 140 K. This value encompasses $J=66$ K
obtained by a fit of a frustrated chain model to the static
susceptibility.\cite{Masuda03}

\begin{figure}[th]
   \begin{center}
    \setlength{\unitlength}{1cm}
    \includegraphics[width=9cm, clip]{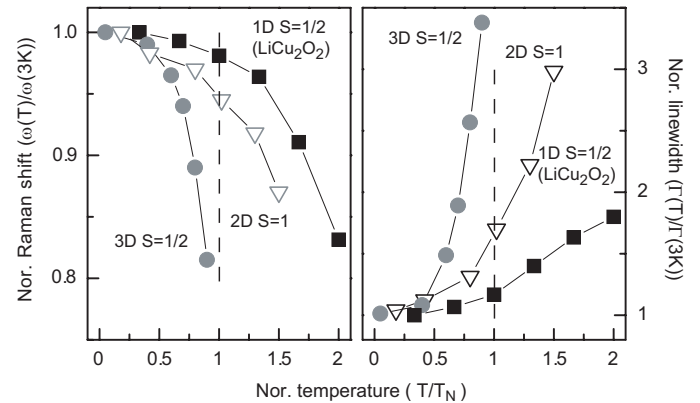}
\end{center}
\caption{ Comparison  of  renormalized frequency (left panel) and
damping (right panel) of two-magnon continuum as a function of
spin number and dimensionality.  The higher dimensional data are
taken from Ref.~16. The dashed vertical line marks a N\'{e}el
temperature.}
\end{figure}
In the following we will discuss the phase II and III.
Quasi-elastic scattering develops in the respective temperature
interval. In addition, a weak continuum extending from 102
cm$^{-1}$ to 120 cm$^{-1}$ becomes clearly distinguishable from
the broad 2M continuum. This continuum persists up to 200 K, that
is, into the dimerized phase III. We thus assign the observed
signal to a 2M continuum which corresponds to a double spin-flip
process of two singlets into a higher singlet
state.\cite{Lemmens03} Since the onset of the continuum
corresponds to twice the spin gap, one obtains a spin gap of
$\Delta\approx 73$ K. This value is in excellent agreement with
$\Delta\approx 72$ K from high-field ESR
measurements.\cite{Zvyagin02}

However, we find no evidence for the presence of additional
magnetic excitations in the magnetic long-range ordered state II.
The silence of magnetic Raman scattering seems to be related to
the incommensurate helimagnetic structure suggested in
Ref.~\cite{Masuda03}. In contrast to the 2M continuum from the
N\'{e}el ordered state, the 2M continuum from singlet states shows
no noticeable change in peak position as well as intensity as a
function of temperature. This is due to the narrow bandwidth of
triplet excitations compared to the magnitude of the spin gap.
Moreover, two 2M continua are simply superposed to each other
independently. This indicates that in phase II short-range singlet
correlations and the helimagnetic ordered state coexist {\it
coherently}. Furthermore, the simultaneous observation of magnetic
excitations from N\'{e}el order and singlet correlations in phase
I can be interpreted as the coherent coexistence of an
antiferromagnetic long-range ordered and a dimerized state which
are mutually exclusive. Note here that the dimerized state is
robust in the course of the evolution of magnetic structures and
their corresponding magnetic correlations. As mentioned above,
such a coexistence has been reported in the impurity-doped
spin-Peierls system CuGeO$_3$~\cite{Regnault95} and the highly
hole-doped quasi-1D cuprate Sr$_{0.73}$CuO$_2$.\cite{Meijer99} The
introduction of nonmagnetic impurities to dimerized states
produces a local moment by breaking up dimers. With aid of higher
dimensional interactions long-range ordering occurs at low
temperature.\cite{Martins97} This disorder-order transition
scenario might be applicable to the studied system as the
dimerized phase of LiCu$_2$O$_2$ is intrinsically contaminated by
a LiCuO impurity phase which has copper ions in a nonmagnetic
Cu$^{+}$ oxidation state. The LiCuO phase is estimated to be less
than $10\%$ of the total volume. It is arranged almost regularly
in the form of platelets with the dimension of $100\times 7\times
100$ nm$^3$ (Ref.\cite{Zvyagin02}). It is thus natural to expect
that similar to a single-site dopant, such nanostructural
nonmagnetic inclusions break Cu$^{2+}$-Cu$^{2+}$ bonding along the
chains, enhance three-dimensional interactions and eventually
promote AFM long-range order at low temperatures.

However, within this mechanism it is difficult to capture the
following features. First, the presence of a helimagnetic state is
not explainable because for the introduction of dopants the
interaction between local moments has an alternating sign. That
is, its sign relies on whether the two local moments belong to the
same or to opposite sublattices. As a result, frustrations are
lifted and an AFM ordering is favored.  Second, the transition of
an incommensurate phase into the N\'{e}el ordered one at 9 K is
driven by the presence of the magnetic
Li$_2$CuO$_2$-impurity.\cite{Masuda03} Third, the 2M signal from a
long-range ordered state is much stronger than that from a
dimerized phase (see Fig.~2). All these observations indicate the
predominance of a long-range ordering over a dimerization as well
as the dependence of the magnetic structure on the kind of
impurities, signalling the significant role of magnetic
impurities. In the following we present a possible scenario.

Upon cooling dilute magnetic impurities which lie most probably
between spin chains undergo long-range ordering  via 3D
interactions. Then, the long-range ordered impurities can be
regarded as effective magnetic fields inducing long-range ordering
in spin chains. The observed helimagnetic state is due to the
compromise of AFM order with strong in-chain frustration.
Furthermore, two successive transitions at 22.5 K and 24.2
K~\cite{Zvyagin02} reflect an intriguing interplay of magnetic
impurities between chains and nonmagnetic in-chain impurities on
the background of the incommensurate spin structure. Upon further
cooling, depending on the presence of the Li$_2$CuO$_2$-impurity,
the N\'{e}el ordered phase will appear at about 9 K. This
indicates that the collinear magnetic structure of the
Li$_2$CuO$_2$-impurity promotes the transition of the
incommensurate phase to the commensurate one.

\begin{figure}[th]
   \begin{center}
    \setlength{\unitlength}{1cm}
    \includegraphics[width= 8cm, clip]{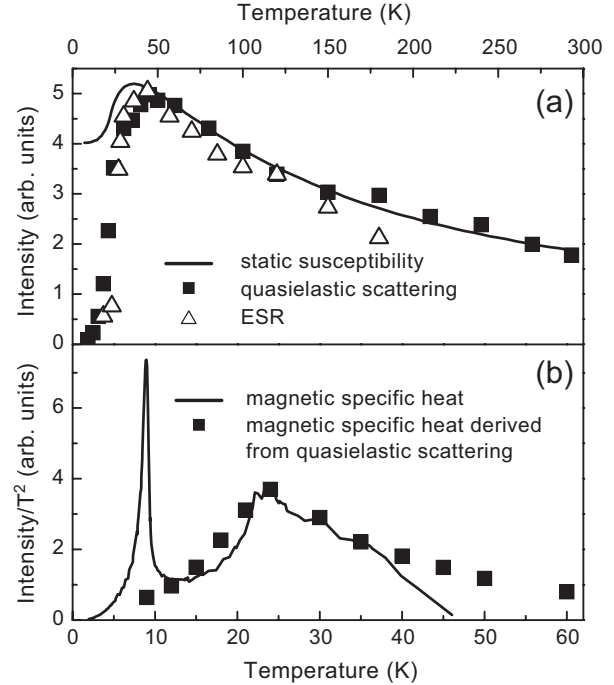}
\end{center}
\caption{ (a) Comparison  of  the intensity of quasielastic
scattering (full rectangle) and ESR (open triangle) to static
susceptibility (solid line) taken from Ref.~10. (b) Mapping of the
magnetic specific heat derived from quasielastic scattering (full
rectangle) on the magnetic specific heat obtained by subtracting
the phonon contribution (solid line).}
\end{figure}

Finally, we will discuss the quasielastic Raman response observed
above 10 K. Our experimental setup is carefully adjusted to
suppress Rayleigh scattering. Therefore, the observed intensity is
intrinsic and origins  from fluctuations of the energy density of
the system. Its presence is a remarkable feature of
low-dimensional system with strong spin-phonon coupling.
~\cite{Lemmens03}

According to the theory of Reiter~\cite{Reiter76} and
Halley~\cite{Halley78}, the scattering intensity is given by the
Fourier component of the correlation function of the magnetic
energy density: $I(\omega)\propto \int^{\infty}_{-\infty} dt
e^{-i\omega t} <E(k,t)E^{\ast}(k,t)>$, where $E(k,t)$ is the
magnetic energy density. In a hydrodynamic assumption for the
correlation function in the high temperature
limit~\cite{Halperin69} the above equation is simplified to the
Lorentzian profile $I(\omega)\propto
\frac{C_{m}T^{2}D_{T}k^{2}}{\omega^{2} +(D_{T}k^{2})^{2}}$, where
$k$ is the scattering wave vector, $D_{T}$ the thermal diffusion
constant, and $C_{m}$ the magnetic specific heat. In this case,
magnetic specific heat is proportional to the integrated intensity
divided by $T^{2}$. Using this relation one can map scattering
intensity on thermodynamic quantities. Figure~4(a) displays
integrated intensity of quasielastic scattering together with ESR
intensity and static susceptibility as a function of temperature.
There exists an overall good correspondence between integrated ESR
and quasielastic intensity. However, they show a substantial
deviation from the static susceptibility below 45 K. Here note
that ESR probes excitations within excited states which originate
from the low-dimensional character of interactions in
LiCu$_2$O$_2$ as the drastic drop of its intensity below 45 K
shows. Thus, one can see that ESR and Raman spectroscopy are
selectively sensitive to disordered short-range correlations. In
contrast, the static susceptibility becomes dominated by 3D
short-range-order antiferromagnetic correlations below 45 K.  In
Fig.~4(b) the magnetic specific heat derived from the quasielastic
scattering is shown together with the magnetic part of specific
heat obtained after subtracting a calculated phonon contribution
from the measured specific heat. In the temperature interval 12 K
$< T <$ 40 K there is a reasonable matching between them. The
difference above 40 K can be attributed to an overestimation of
the phonon contribution to the specific heat at high temperatures
by choosing only one Debye function. The discrepancy below 12 K
comes from the suppression of fluctuations of spin energy density
in the N\'{e}el ordered phase.

A maximum of $C_{m}$ can also provide an information on the
exchange constant. In the case of 1D AF chain the magnetic
specific heat has a broad maximum at $k_{B}T\approx 0.481 J$. The
frustration shifts the maximum of $C_{m}$ to lower temperature,
$k_{B}T \sim 0.38 J$.~\cite{Kuroe97} This results in $J\sim 60$ K
which is consistent with the value obtained from the peak position
of the 2M scattering as well as from the static susceptibility.

In summary, a Raman scattering study of the frustrated spin chain
compound LiCu$_{2}$O$_{2}$ unveils a coexistence of dimerized
correlations and long-range order as a coherent superimposition of
a dimerized state with a spin gap $\Delta\sim 73$ K on a
two-magnon continuum from a N\'{e}el ordered state shows. This is
attributed to nano-scale sized nonmagnetic defects. Moreover, the
transition of a helimagnetic state into a N\'{e}el ordered one
suggests a significant role of magnetic impurities in an
incommensurate magnetic structure as a driving impetus to a
commensurate structure. We hope that our experimental observations
will stimulate further theoretical investigations devoted to
understanding magnetic properties of frustrated spin chain systems
including magnetic/nonmangetic defects.

The work  was supported by DFG/SPP 1073, INTAS 01-278 and NATO
Collaborative Linkage Grant PST.CLG.977766.


\begin{thebibliography}{99}
\bibitem{helim}
Strictly speaking, the phase transition into a helimagnetic phase
around 23 K consists of two successive transitions at 22.5 K and
24.2 K~\cite{Zvyagin02}. The possible origin will be discussed
below.
\end{thebibliography}
\end{document}